\documentclass[a4paper,11pt]{article}

\usepackage{jinstpub}

\title{A portable muon telescope for multidisciplinary applications}

 \author[a,1]{R.M.I.D Gamage, \note{Corresponding author}}
 \author[a]{S. Basnet,}
 \author[a]{E. Cortina Gil,}
 \author[a]{P. Demin,}  
 \author[a]{A. Giammanco,}
 \author[a]{R. Karnam,}
 \author[a]{M. Moussawi,}
 \author[b]{M. Tytgat}

 \affiliation[a]{Centre for Cosmology, Particle Physics and Phenomenology (CP3),\\Universit\'e catholique de Louvain,\\Louvain-la-Neuve, Belgium}
 
 \affiliation[b]{Department of Physics and Astronomy,\\Ghent University,\\Ghent, Belgium}

\emailAdd{ishan.ran@uclouvain.be}

\abstract{Muon tomography or “muography” is an emerging imaging technique that uses cosmogenic muons as the radiation source. Due to its diverse range of applications and the use of natural radiation, muography is being applied across many fields such as geology, archaeology, civil engineering, nuclear reactor monitoring, nuclear waste characterization, underground surveys, etc. Muons can be detected using various detector technologies, among which, resistive plate chambers (RPC) are a very cost effective choice. RPCs are planar detectors which use ionization in a thin gas gap to detect cosmic muons, already used since years in major particle accelerator experiments.

We have developed a muon telescope (or “muoscope”) composed of small scale RPCs. The design goal for our muoscope is to be portable and autonomous, in order to take data in places that are not easily accessible. The whole setup is light and compact, such to be easily packed in a car trunk. Individual RPCs are hosted in gas-tight aluminium cases. There is no need for gas bottles, once the chambers are filled. The muoscope can be controlled from a reasonable distance using wireless connection. 

In this paper we summarize the guiding principles of our project and present some recent developments and future prospects, including a long-term stability study of the resistivity of the semiconductive coating obtained with serigraphy.
}


\proceeding{XXII International Workshop on Radiation Imaging Detectors - iWoRiD 2021 \\27 June- 1 July 2021 \\Ghent, Belgium}

\begin{document}
\maketitle
\flushbottom

\section{Introduction}

Muography (i.e., muon radiography or muon tomography) is an imaging technique that uses high energetic cosmic muons, as a radiation source. Conceptually, it is similar to common radiography, but instead of X-ray photons other elementary particles, the muons, are used.  Muons are produced when primary cosmic rays interacts with the upper earth atmosphere, and the flux at sea level can be used in the imaging of a target volume. This imaging technique determines the density of a structure of interest by measuring the flux of muon passing through the unobstructed sky or through a known volume of material.

Muography has several applications in different fields~\cite{Bonechi:2019ckl}. In the last decade, large muon telescopes have been developed for applications e.g. in volcanology, where the size of the object of interest requires the detectors to be at large distance. Muography also finds its applications in the imaging of smaller targets, such as nuclear waste monitoring, civil engineering, archaeology, etc.; in some cases, the optimal view point may be located in narrow and confined environments like tunnels or underground chambers. The first application of muography dates back to the middle of the past century~\cite{George1955}, to measure the overburden of a tunnel. More than 20 years later, Malmqvist et al.~\cite{Malmqvist1979} studied the sensitivity of the method to mineral exploration, and addressed some of the constraints coming from the availability of tunnels of suitable size and location to host particle detectors. Mineral exploration with muography is nowadays in a stage of relative maturity, and already in use for commercial purposes by dedicated companies~\cite{Lingacom2018,Schouten2018}. Some teams are already developing portable detectors for muography, typically based on scintillator bars~\cite{Baccani:2018nrn} or fibers~\cite{Kyushu2018}.

We are developing a Resistive Plate Chamber (RPC) based muon telescope (or "muoscope") with a relatively small scale active area. An RPC is a gaseous detector with parallel plate geometry~\cite{Santonico-Cardarelli_1981,AbbresciaPeskovFonte_2018}. It consists of two highly resistive plates which are placed parallel to each other, with semi-conducting coating on the exterior sides of the plates. The volume between the two glasses is filled with an appropriate gas mixture, and high voltage is supplied to the two resistive layers.
RPCs are popular in particle physics experiments~\cite{AbbresciaPeskovFonte_2018}, and have already been used in a few muography applications~\cite{Menedeu:2016rrr,Baesso_2013,TUMUTY_pan_2019}, thanks to their favourable trade-off between cost and performance~\cite{MuographyBook}; they tend to be used for large-area detectors.
The novelty of our RPC-based muoscope is instead to be very compact. The design goal of developing small-area RPC detectors that are also portable and autonomous brings new challenges, that we describe in this paper.

\section{Guiding principles of the project}
\label{sec:principles}

We aim at developing a portable muon telescope that can be used for applications where confined space is available to deploy the detectors, e.g. speleology, archaeometry, mineral searches, as well as the imaging of cultural heritage, such as monumental statues and building decorations. For the latter type of applications, imaging methods based on other forms of radiation (X rays, neutrons) have been used with success in some cases~\cite{morigi2010application}, but become infeasible when the objects of interest are too large to be transported to a properly equipped laboratory. 

In this context, our goals include portability, robustness, safety and autonomy. 
Portability imposes strong limitations in both weight and size; robustness means the ability to operate in a wide range of ambient conditions; safety is particularly important in confined environments, and it is not trivial for gaseous detectors; and autonomy means that the detectors should run for extended periods without human intervention, for example by using a battery-based, very low-power readout and control electronics able to adapt to changing environmental conditions. 

\begin{figure}[htbp]
\centering
\includegraphics[width=0.40\textwidth]{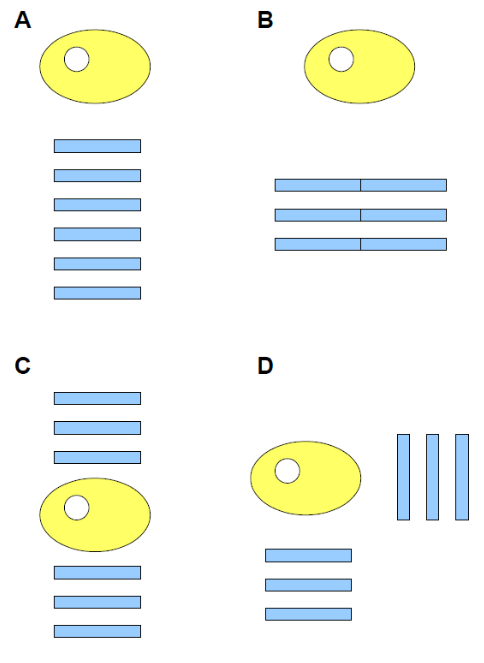}
\caption{\label{fig:versatility} Examples of detector configurations among which one could switch with fast turn-around. The same six RPCs may be deployed such to maximize angular resolution and trigger
redundancy (A) or to maximize the acceptance and therefore the
statistics (B). If two tunnels are available at appropriate locations,
the same RPCs can be used to "sandwich" the volume of interest (C)
or to measure projections from orthogonal points of view (D).}
\end{figure}

Similarly to many other gaseous detectors, RPCs are typically operated with continuous gas flow through the detector using external gas supply (mainly through attached gas bottles), and also uses some dangerous gas mixtures (toxic and possibly flammable) in order to operate with desired efficiencies and modes. 
The problem of reducing the necessary gas flow has been studied, for example, in Refs.~\cite{procureur2020we,Assis:2020mvp,nyitrai2021toward}.
In our project, we are using small scale (active area of $16\times 16~cm^2$) air-tight gas chambers, allowing a stable operation during several weeks. Not needing gas bottles or gas flow (gas is filled into the chamber before starting operations) allows us to use the muoscope in confined areas and also facilitates portability, by reducing weight and size of the overall setup.
An additional design consideration of our project is modularity, and therefore versatility, of the setup. Each RPC is housed in an individual aluminium casing, and we foresee to keep them separated by spacers, in arrangements that can be adapted to various use cases. For example, four RPCs can be arranged in two adjacent pairs, alternating in orthogonal orientations, such that even and odd RPCs measure, respectively, the X and Y coordinates of the point of passage of the muon. Spacers, in this case, are only used to separate the second and third RPC. This way, the full X-Y information is available at two distinct positions along the Z axis, allowing the reconstruction of a 3D muon trajectory. However, if one is only interested in a 2D projection of the muon trajectory, the same four layers can be disposed with parallel orientation, all of them equally spaced such that there are four 1D positions measured along the Z-axis. Such alternative configuration would give a more reliable track selection (by trigger redundancy) and a better angular precision, although limited to only one angle. Of course, many more combinations become possible once more RPCs are produced, as sketched for the example of six RPCs in Figure~\ref{fig:versatility}. Differently from other portable muon telescopes developed elsewhere, for which the setup is fixed, in our case thanks to the modularity of the setup, a user can switch between configurations with no effort.

\section{Technical description of the current prototype}
\label{sec:techdescription}

The RPCs are gaseous detectors, i.e. the active detecting medium is a gas mixture, where the optimized gas mixture is filled in the gap between two highly resistive parallel plates (glass in our muoscope). 
The exterior sides of these two plates are painted with a semi-conductive coating that allows to spread high voltage (HV) throughout the plate and makes a uniform electric field across the gas volume. When a charged particle passes through the gas gap, it ionizes the gas molecules along its path, releasing free electrons that accelerate toward the electrode, collide with gas molecules and produce an avalanche. The motion of these charges induces a signal in the metallic readout strips which are read individually, providing position information on the deposited charge. 

\begin{figure}[htbp]
\centering
(a)
\includegraphics[width=0.4\textwidth]{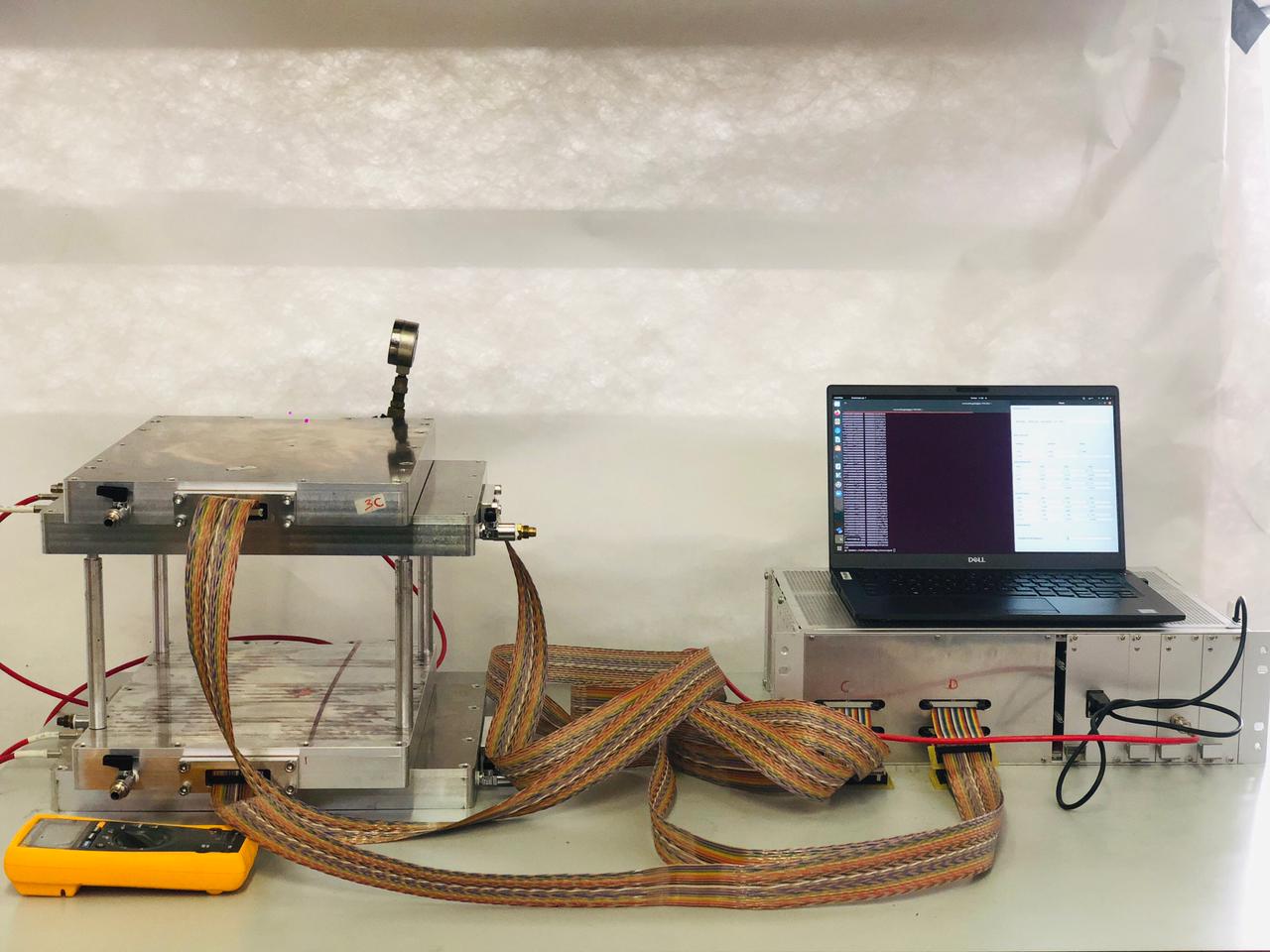}
(b)
\includegraphics[width=0.40\textwidth]{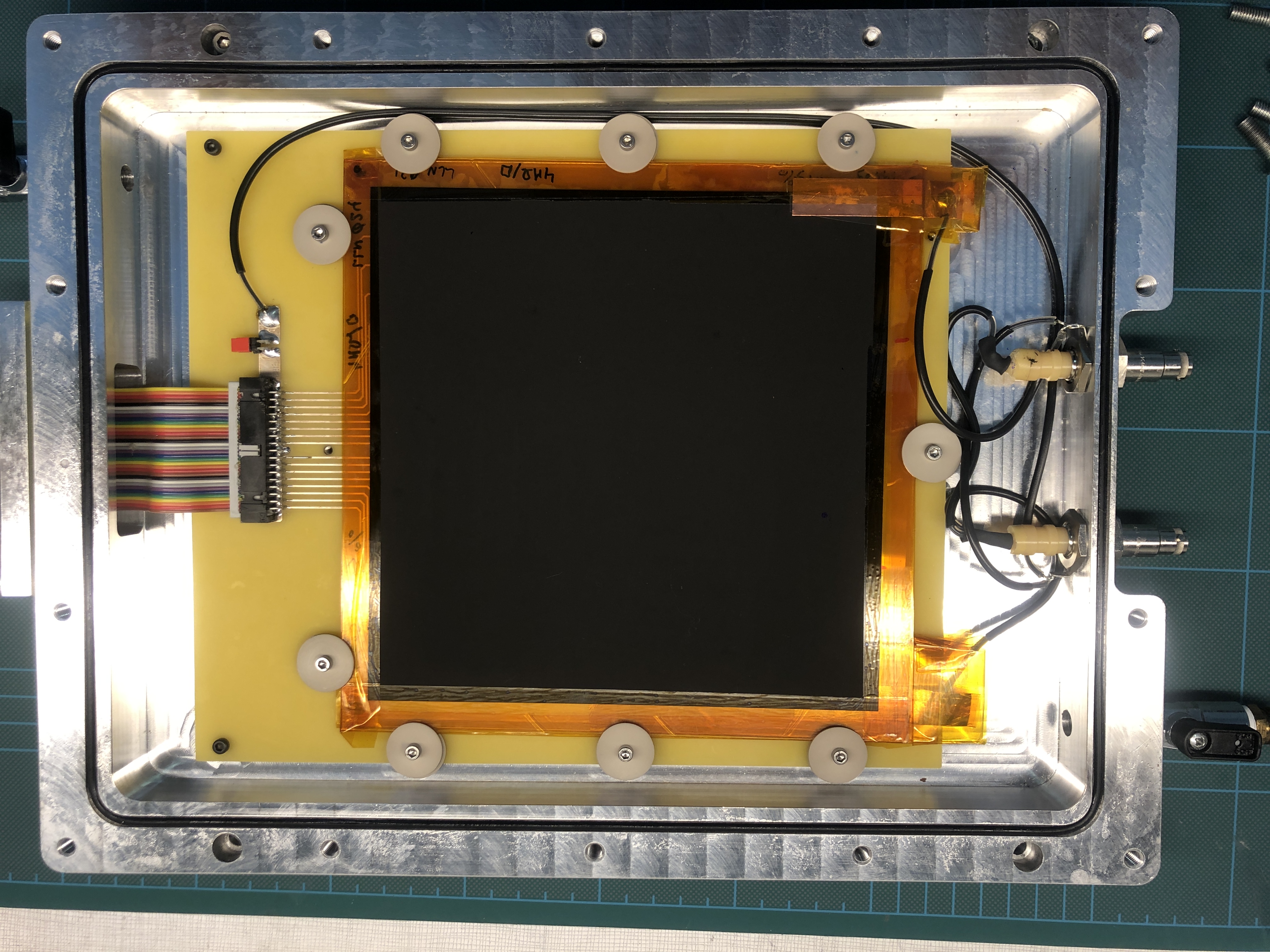}
\caption{\label{fig:detector} (a) Our muoscope with 2 RPC detectors.  (b) One of the RPC detectors inside its casing.}
\end{figure}

The current muoscope prototype consists of four identical RPCs. The data acquisition system (DAQ) is integrated with the HV supply to the RPCs. In the example configuration shown in Figure~\ref{fig:detector}(a), the first and third RPCs are oriented orthogonal to the second and fourth, in order to provide a bi-dimensional information (X and Y orientation).


Each RPC, shown in Fig.~\ref{fig:detector}(b), is built with 1~mm thick glass plates and an air gap of 1.1~mm. A uniform distance between the glass plates is obtained with nine round edge spacers made of polyether ether ketone (PEEK). 
Sixteen copper strips are used for the readout of RPC signals. Each strip is 0.9~cm wide and separated by a 0.1~mm gap from the adjacent strips, making a pitch of 1~cm.
Each RPC is hosted in an air-tight aluminum box. The rate of gas leakage in vacuum conditions was measured using helium to be 10$^{-9}$ mbar~l~s$^{-1}$~\cite{Wuyckens2018}.
Each detector layer (casing + RPC) weighs 6.5 kg. The current gas mixture consists of R134a Freon (95.2\%), isobutane (4.5\%) and SF$_{6}$ (0.3\%), kept at a pressure slightly above (by $\sim$0.1 atm) the atmospheric one.
The data-acquisition electronics consist of two front-end boards (FEBs) borrowed from the Muon RPC system of the CMS experiment~\cite{FEB1, FEB2}. Each FEB can handle 32 analog inputs channels, each consisting of an amplifier with a charge sensitivity of 2 mV/fC, a discriminator, a monostable and a LVDS driver. The LVDS outputs of all the FEBs are connected to a System-on-Chip (SoC) module, which is installed on a carrier board with a wireless connection, to ensure autonomy.

\section{Semi-conductive coating by serigraphy}
\label{sec:serigraphy}

In the first prototype~\cite{Wuyckens2018}, the semiconductive coating of the glass sheets was laid by hand with a paint roller. 
This procedure, in addition to not being sustainable for large numbers of RPCs, was found to lead to non-uniformities of up to $200\%$.
This motivated our interest for serigraphy.

\begin{figure}[htbp]
\centering
(a)
\includegraphics[width=0.23\textwidth,height=0.32\textwidth]{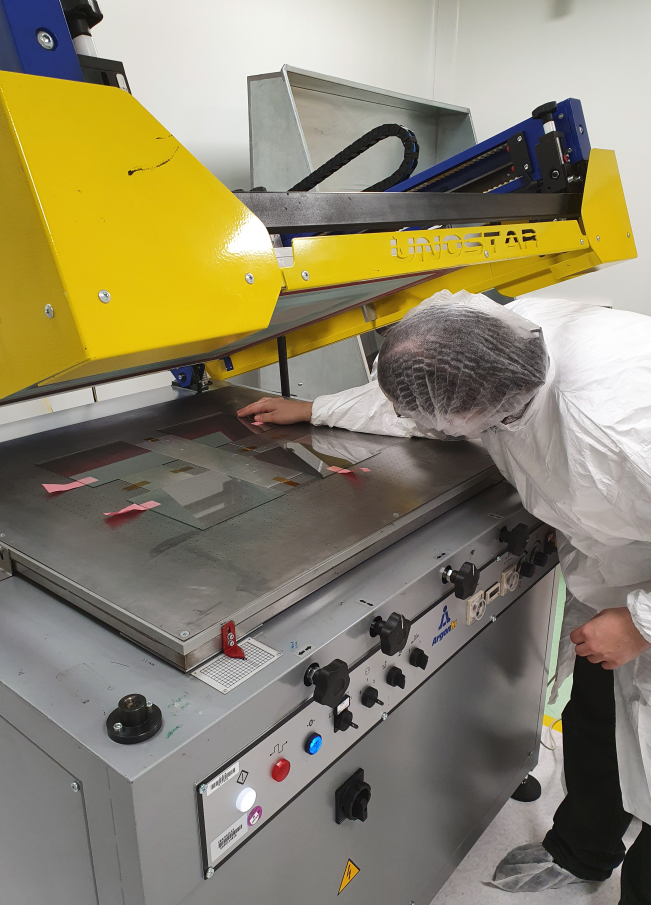}
(b)
\includegraphics[width=0.23\textwidth,height=0.32\textwidth]{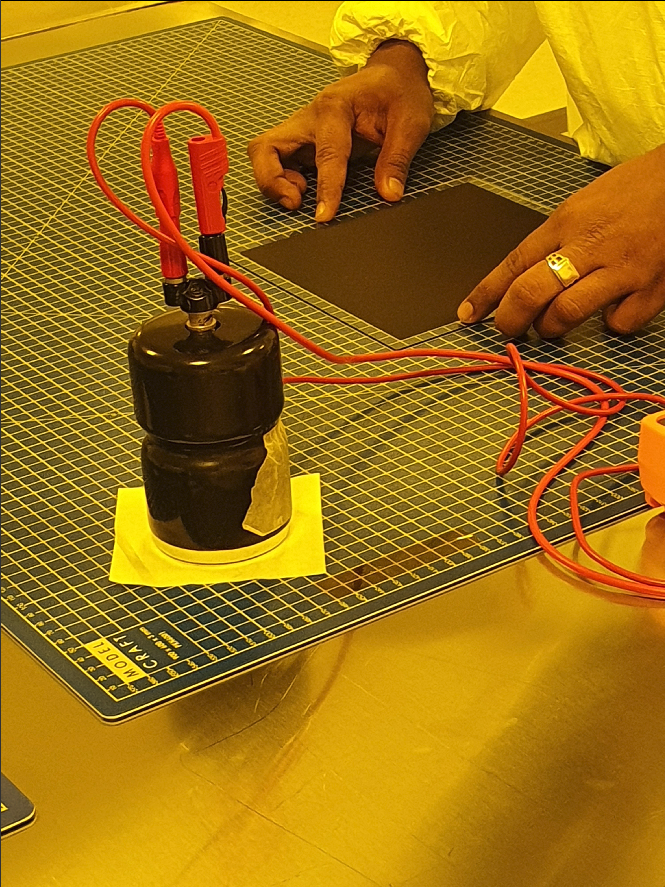}
(c)
\includegraphics[width=0.25\textwidth,height=0.25\textwidth]{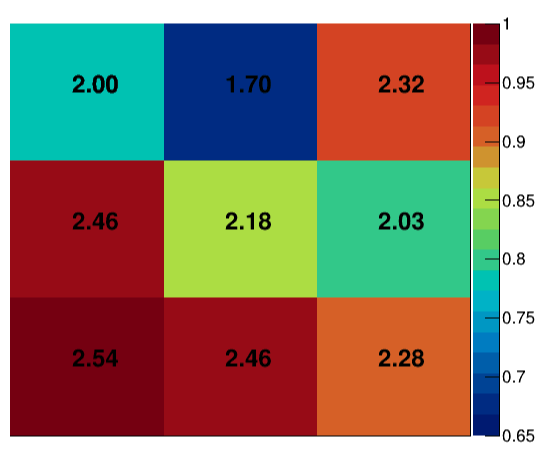}
\caption{\label{fig:serigraphy}(a) Serigraphy machine at CEA-Saclay, preparing for printing.  (b) Printed glass plate, preparing for resistivity measurement. (c) Surface resistivity as measured in February 2020 in nine different regions of a representative glass plate.}
\end{figure}

Serigraphy, also known as screen-printing, is one of the oldest printing techniques and consists in using a fine mesh to transfer ink onto a surface. Similarly, dedicated serigraphy machines, as in Fig.~\ref{fig:serigraphy}(a), can be used to cover glass plates with a semi-conductive paste. 
In February 2020, in preparation for our next prototype, we procured glass plates with uniform resistive layers using the equipment and expertise available at CEA in Saclay, France. 
Details about the composition and preparation of the semi-conductive paste used for our new glasses have been provided in a previous publication~\cite{Basnet2020}, where we also reported that the non-uniformity went down to about 20\%. In the rest of this section, we report on our long-term studies of the evolution of the resistivity and of its uniformity with time, a rather important consideration for our project.


\begin{figure}[htbp]
\centering
\includegraphics[width=1\textwidth]{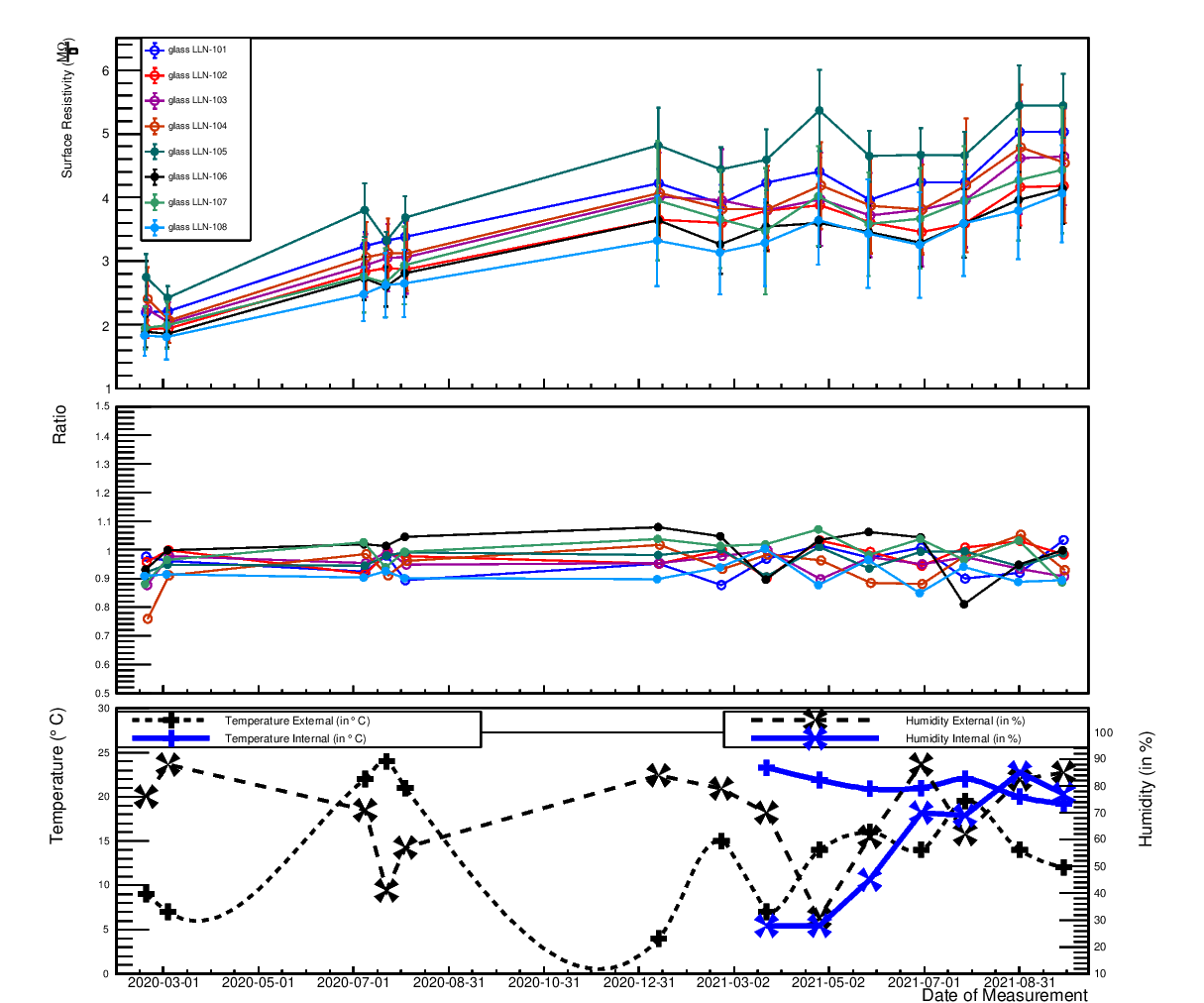}
\caption{\label{fig:resistivity}
Evolution with time of the following observables of interest.
Top: average surface resistivity of eight selected glass plates. The vertical bars represent the standard deviations of the measurements at nine different locations on the same plate.
Middle: ratio, for each of the same glass plates, between the resistivity at the center and the average resistivity of the other eight locations. 
Bottom: temperature and humidity as measured externally (i.e. data from the local weather casting) and internally (i.e. data from an internal Arduino-based sensor).
}
\end{figure}

For each glass plate, we measure the surface resistivity at nine different spots, see Figure~\ref{fig:serigraphy}(b); a sample measurement for a single plate is shown in Figure~\ref{fig:serigraphy}(c).
Being interested in long-term variations  of surface resistivity, eight selected glass plates are being measured since February 2020 in order to check the stability of the resistivity with the time and with respect to the environmental conditions, i.e. temperature and humidity. We present out data in figure \ref{fig:resistivity}. 
The large gaps in the time axis are due to Covid-19 pandemic lockdown in Belgium, that limited the access to our laboratory. 
The top panel shows, for each glass plate and for each time, the average and standard deviation of the nine different measurements. A clear upward trend is visible with time.
Given our interest in spatial uniformity, in the middle panel we show the ratio of the surface resistivity of the central region of each plate to the average of the other eight regions of Figure~\ref{fig:serigraphy}(c). 
Unlike the increasing trend in the overall surface resistivity, the fact that these ratios have been more or less constant so far is rather reassuring, as it indicates that non-uniformity does not increase with time. 
Finally, in the bottom panel, we show the humidity and temperature at the time of resistivity measurement. 
Initially we could only take the temperature and humidity from the local weather casting (``external'' measurements, in black) and then, for the last six months, we were able to measure the temperature and humidity using an Arduino-based sensor during the resistivity measurements itself (``internal'' measurements, in blue). For the next measurements, we will add data from a new sensor unit that records the temperature and humidity every day.

Although no obvious correlation has been observed so far between resistivity and environmental variables, we plan to continue taking these measurements. In case any correlation is found, we believe we can use these results to {\it ex situ} correct for the surface resistivity once the glass plates are installed inside the chambers and the muoscope is out for data-taking.

\section{Outlook}
\label{sec:future}

The main goal of the first prototype was to acquire experience with a complete detection system that fulfills the basic requirements of compactness and autonomy.  
This section presents and motivates some of the groundwork that is now ongoing in preparation for the next prototypes.


The current prototype is operated in self-triggered mode. While this is expected to be the normal mode of operation, especially in the field, there are cases where it is convenient to switch to a reliable independent trigger from an external source (e.g. for performance studies). 
For this motivation, we recently built and validated an external trigger system based on plastic scintillators. As the next step, we are designing a new printed circuit board to integrate the signals from this external trigger system into the main DAQ.

Our initial choice of aluminum as material for the cases that host the RPCs may be replaced by carbon fibers or other light materials compatible with 3D printing. 
That would yield a much lighter detection setup, and with less scattering of the low energy muons; however, a thorough dedicated study will be needed to quantify the possible contamination of the operating gas due to outgassing from the casing itself~\cite{procureur2020we}, which is a major concern given the requirement of no constant flushing.


Some of the parameters of the muoscope were initially chosen based on convenience, rather than optimization. 
This includes the strip width, currently much larger than the intrinsic resolution that is known to be achievable with RPCs~\cite{MuographyBook}, and the number of strips, which was dictated by the available read-out electronics at the start of this project. 
High-resolution imaging would require a significant increase in the number of channels, for which we are in the process of moving to the MAROC ASIC \cite{Barrillon:1091460}. 
We use the Garfield++ open-source code~\cite{Garfield} to simulate signal formation, in view of optimizing RPC parameters such as the distance between strips and the gap between the glasses. 
We are also studying the possibility to use pixels instead of strips; however, the corresponding increase in the number of read-out channels means also an increase in overall cost, and in power consumption. For this reason we are also working on new multiplexing approaches in order to reduce the total amount of electronic channels by a large  factor  without compromising in position resolution.

\acknowledgments

This work was partially supported by the EU Horizon 2020 Research and Innovation Programme under the Marie Sklodowska-Curie Grant Agreement No. 822185, and by the Fonds de la Recherche Scientifique - FNRS under Grants No. T.0099.19 and J.0070.21. Samip Basnet and Raveendrababu Karnam would like to acknowledge additional research grants from FNRS (FRIA doctoral grant) and UCLouvain (FSR postdoctoral fellowship), respectively. 
We thank Dr. Stephan Aune and his serigraphy team at CEA, Saclay for helping us with resitive coating. 
We gladly acknowledge Sophie Wuycken's contribution at the start of this project. 
We also thank the electronics group at the Center for Cosmology, Particle Physics, and Phenomenology (CP3), Universit\'{e} catholique de Louvain, for their help in setting up the external scintillator trigger.


\bibliographystyle{unsrt}
\bibliography{refs}

\end{document}